\documentclass[showpacs,twocolumn,JCP,superscriptaddress]{revtex4}
\usepackage{graphicx}
\usepackage{amssymb}
\usepackage{amsmath}
\usepackage{amsfonts}
\usepackage{mathrsfs}

\renewcommand{\epsilon}{\varepsilon}
\newcommand{\figurewidth}{0.45\textwidth}

\begin{document}
\title{Dynamics of DNA translocation through an attractive nanopore}

\author{Kaifu Luo}
\altaffiliation[]{
Author to whom the correspondence should be addressed}
\email{luokaifu@gmail.com}
\affiliation{Department of Engineering Physics, Helsinki University of Technology,
P.O. Box 1100, FIN-02015 TKK, Espoo, Finland}
\affiliation{Physics Department, Technical University of Munich,
D-85748 Garching, Germany}
\author{Tapio Ala-Nissila}
\affiliation{Department of Engineering Physics, Helsinki University of Technology,
P.O. Box 1100, FIN-02015 TKK, Espoo, Finland}
\affiliation{Department of Physics, Box 1843, Brown University, Providence,
Rhode Island 02912-1843, USA}
\author{See-Chen Ying}
\affiliation{Department of Physics, Box 1843, Brown University, Providence, Rhode Island
02912-1843, USA}
\author{Aniket Bhattacharya}
\affiliation{Department of Physics, University of Central Florida, Orlando, Florida
32816-2385, USA}

\date{August 19, 2008}

\begin{abstract}

We investigate the dynamics of DNA translocation through a nanopore
driven by an external force using Langevin dynamics simulations in
two dimensions (2D) to study how the translocation dynamics depend
on the details of the DNA sequences. We consider a coarse-grained
model of DNA built from two bases $A$ and $C$, having different
base-pore interactions, {\textit e.g.}, a strong (weak) attractive
force between the pore and the base $A$ ($C$) inside the pore.
From a series of studies on hetero-DNAs with repeat units $A_mC_n$,
we find that the translocation time decreases exponentially as a
function of the volume fraction $f_C$ of the base $C$. 
For longer $A$ sequences with $f_C \le 0.5$, the translocation time 
strongly depends on the orientation of DNA, namely which base enters the pore first.
Our studies clearly demonstrate that for a DNA of certain length $N$ with repeat 
units $A_mC_n$, the pattern exhibited by the waiting times of the individual bases 
and their periodicity can unambiguously determine the values of $m$, $n$ and $N$ 
respectively. Therefore, a prospective experimental realization of this phenomenon 
may lead to fast and efficient sequence detection technic.   

\end{abstract}


\maketitle
\section{Introduction}

Polymer translocation through a nanopore is a challenging problem in polymer physics
and it also plays a critical role in numerous biological processes, such as DNA and
RNA translocation across nuclear pores, protein transport through membrane channels,
and virus injection.

In a seminal experimental paper, Kasianowicz \textit{et
al.}~\cite{Kasianowicz} demonstrated that an electric field can
drive single-stranded DNA and RNA molecules through the water-filled
$\alpha$-hemolysin channel and that the passage of each molecule is
signaled by a blockade in the channel current. These observations
can be directly used to characterize the polymer length. Triggered
by their experiments and various potential technological
applications~\cite{Kasianowicz,Meller03},
such as rapid DNA sequencing, gene therapy and controlled drug delivery, 
the polymer translocation has become a subject of intensive experimental
~\cite{Akeson,Meller00,Meller01,
Meller02,Meller07,Bashir,Sauer,Mathe,Henrickson,Li01,Li03,Li05,Krasilnikov,Keyser1,
Keyser2,Dekker,Trepagnier,Storm03,Storm05,Storm052}
and theoretical
~\cite{Storm05,Storm052,Simon,Sung,Park,diMarzio,Muthukumar99,MuthuKumar03,Lubensky,
Kafri,Slonkina,Ambj,Metzler,Ambj2,Ambj3,Baumg,Chuang,Kantor,Panja,Dubbledam,Milchev,Luo1,Luo2,
Luo3,Luo4,Luo5,Huopaniemi1,Huopaniemi2,Chern,Loebl,Randel,Lansac,Kong,Farkas,
Tian,Liao,Zandi,Kotsev,Tsuchiya,Bockelmann}
studies.

One of the fundamental questions that has been addressed in the
community is how the translocation time $\tau$ depends on the system
parameters and the polymer characteristics, such as: chain length
$N$~\cite{Meller01,Meller02,Storm05,Storm052,Sung,Muthukumar99,Lubensky,
Chuang,Kantor,Panja,Dubbledam,Milchev,Luo1,Luo2,Luo3,Luo4,Huopaniemi2,Huopaniemi1,Tian},
sequence and secondary
structure~\cite{Akeson,Meller00,Meller02,Sauer,Kafri,Luo3,Luo4,Luo5,Kotsev,Tsuchiya,Bockelmann},
pore length $L$ and width $W$~\cite{Luo1}, driving force
$F$~\cite{Meller01,Meller02,Henrickson,Sauer,
Kantor,Luo2,Luo3,Luo4,Huopaniemi1,Huopaniemi2,Chern,Loebl,Tian}, and
polymer-pore
interactions~\cite{Meller00,Meller02,Krasilnikov,Lubensky,Luo4,Luo5,Tian}.
A central issue from the point of view of sequencing is whether DNA
translocation through a nanopore can be used to determine the
detailed sequence structure of the molecule~\cite{Luo5}.

It has been demonstrated
experimentally~\cite{Meller00,Meller02,Krasilnikov} that DNA
translocation dynamics is strongly influenced by {\em
nucleotide-pore} interactions. 
Intuitively, for sufficiently strong attraction, the residence time
of each monomer in the pore should increase, resulting in a much
longer translocation time. In a recent Letter, we have investigated
the influence of base-pore interaction on the translocation dynamics
using LD simulations~\cite{Luo4}. The results show that an
attractive interaction increases translocation time slowly for weak
attraction while an exponential increase is observed for strong
attraction in the activated regime~\cite{Tian}. Under weak driving
force and strong attractive force, the translocation time shows
non-monotonic behavior. Our results are in good agreement with
experimental findings~\cite{Meller00,Meller02,Krasilnikov}.

While a DNA is composed of four different
nucleotides~\cite{Alberts}, to date most of the theoretical
treatments have focused on scaling and universal aspects of
translocation of a homogeneous polymers, although several
experiments~\cite{Sauer,Meller03} show that in the real biological
systems inhomogeneities in the structure and interactions between
polymer and other molecules might have a significant effect on the
overall dynamics. Previously, we have considered heteropolymers
consisting of two types of monomers labeled $A$ and $C$, which are
distinguished by the magnitude of the driving force that they
experience inside the pore~\cite{Luo3}. This model captured some
essential features of the heteropolymer translocation. However, for
real biopolymers such as DNA and RNA, no charge difference between
the monomers (nucleotides) exists. Instead, the nucleotide-pore
interactions are base specific~\cite{Meller00,Meller02}. Thus a more
realistic  model for  studying the influence of structure on
biopolymers translocation should differentiate bases $A$ and $C$ by
different base-pore interactions. In a recent letter~\cite{Luo5}, we
have adopted such a  model and investigated hetero-DNAs with
symmetric blocks using Langevin dynamics simulations. We found that
the translocation time depends strongly on the block length as well
as on the orientation of which base entering the pore first.

In this paper we extend our results for the DNA composed of symmetric blocks,   
furnish more general characteristics of the residence time inside the pore, and provide 
more comprehensive results for the translocation
dynamics of various sequences of a two component hetero-DNA. 
We also provide a simple interpretation for the
sequence dependence of the monomer waiting time distribution 
and the total translocation time. The paper is organized as follows.
In section II, we briefly describe our model and the simulation
technique. In section III, we present our results. Finally, the
conclusions and discussion are presented in section IV.

\section{Model and method} \label{chap-model}
In our numerical simulations, the polymer chains are modeled as
bead-spring chains. Excluded volume interaction between monomers is
modeled by a short range repulsive LJ potential: $U_{LJ}
(r)=4\epsilon [{(\frac{\sigma}{r})}^{12}-{(\frac{\sigma}
{r})}^6]+\epsilon$ for $r\le 2^{1/6}\sigma$ and 0 for
$r>2^{1/6}\sigma$. Here, $\sigma$ is the diameter of a monomer, and
$\epsilon$ is the depth of the potential. The connectivity between
neighboring monomers is modeled as a Finite Extension Nonlinear
Elastic (FENE) spring with $U_{FENE}
(r)=-\frac{1}{2}kR_0^2\ln(1-r^2/R_0^2)$, where $r$ is the distance
between consecutive monomers, $k$ is the spring constant and $R_0$
is the maximum allowed separation between connected monomers.

\begin{figure}
  \includegraphics*[width=\figurewidth]{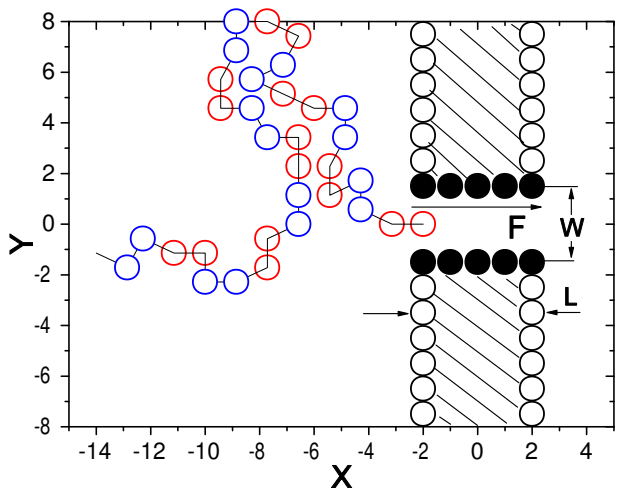}
\caption{ A schematic representation of the system. The pore length
$L=5$ and the pore width $W=3$.
        }
 \label{Fig1}
\end{figure}

We consider a 2D geometry as shown in Fig. \ref{Fig1}, where the
walls along the $y$ direction are formed by stationary particles
within a distance $\sigma$ from each other. The pore of length $L$
and width $W$ are formed from two rows of stationary particles
represented by black circles in Fig. \ref{Fig1}. Between all
monomer-wall particle pairs, there exists the same short range
repulsive LJ interaction as described above. The base-pore
interaction is modeled by a LJ potential with a cutoff of
$2.5\sigma$ and interaction strength $\epsilon_{pA}$ for the base
$A$ and $\epsilon_{pC}$ for the base $C$. This interaction can be
either attractive or repulsive depending on the position of the
monomer from the pore particles.

In the Langevin dynamics simulation, each monomer is subjected to
conservative, frictional, and random forces, respectively,
with~\cite{Allen} $m{\bf \ddot
{r}}_i =-{\bf \nabla}({U}_{LJ}+{U}_{FENE})+{\bf F}_{\textrm{ext}}
-\xi {\bf v}_i + {\bf F}_i^R$, where $m$ is the monomer's mass,
$\xi$ is the friction coefficient, ${\bf v}_i$ is the monomer's
velocity, and ${\bf F}_i^R$ is the random force which satisfies the
fluctuation-dissipation theorem.
The external force is expressed as ${\bf F}_{\textrm{ext}}=F\hat{x}$,
where $F$ is the external force strength exerted on the monomers in
the pore, and $\hat{x}$ is a unit vector in the direction along the
pore axis.

The LJ parameters $\epsilon$, $\sigma$ and the bead mass $m$ fix the
system energy, length and mass units respectively, leading to the
corresponding time scale $t_{LJ}=(m\sigma^2/\epsilon)^{1/2}$ and
force scale $\epsilon/\sigma$. In our model, each bead corresponds
to a Kuhn length of a single-stranded DNA containing approximately
three nucleotide bases, so the value of $\sigma \approx 1.5$
nm~\cite{Kuhn_length}. The average mass of a base in DNA is about
312 amu, so the bead mass $m \approx 936$ amu. We set
$k_{B}T=1.2\epsilon$, so the interaction strength $\epsilon$
corresponds to $3.39 \times 10^{-21}$ J at a temperature 295 K. This
leads to a time scale of 32.1 ps and a force scale of 2.3 pN. The
remaining dimensionless parameters in the model are chosen to be
$R_0=2$, $k=7$, $\xi=0.7$. In addition, the driving force for a bead
in the pore is set as $F=0.5$ unless otherwise stated. This value
corresponds to a voltage of about 187.9 mV across the pore (assuming
three unit charges on a bead and the effective charge 0.094e for a
unit charge~\cite{Sauer,Mathe}), within the range of experimental
parameters~\cite{Kasianowicz,Meller00,Meller01,Meller02,Meller03}.
The pore width $W=3$. This ensures that monomers $A$ and $C$
encounter an attractive force inside the pore~\cite{Note}. The
base-pore interactions $\epsilon_{pA}=3.0$ and
$\epsilon_{pC}=1.0$ are chosen based on comparison of the
theoretical results~\cite{Luo4} with the experimental
data~\cite{Meller00} for the translocation time distribution
histogram of poly(dC)$_{100}$ and poly(dA)$_{100}$. It is worthwhile
to note that the translocation time depends strongly on the pore
length $L$. We have checked that $L \approx 10$ nm produces average
translocation times $\tau \approx $100 $\mu$s in accordance with the
experimental data \cite{Meller00}. For computational efficiency we
present results for $L=5$ ($7.5$ nm) here. The Langevin equation is
integrated in time by a method described by Ermak and
Buckholz~\cite{Ermak} in 2D.

To create the initial configuration, the first monomer of the chain
is placed in the entrance of the pore. The polymer is then allowed
to relax to obtain an equilibrium configuration.
The translocation time is defined as the time interval between the
entrance of the first bead into the pore and the exit of the last
bead. The estimate for the translocation time was obtained by
neglecting any failed translocation and then calculating the average
duration of the successful translocation. Typically, we average our
data over 2000 independent runs with different initial conditions.

\section{Results and Discussions}

\subsection{Recognizing position-specific base $A$ substitutions in poly(dC)}

As demonstrated by experiments~\cite{Kasianowicz,Meller03}, a
nanopore might be used for DNA sequencing. The most fundamental
concern has been whether $\alpha$-hemolysin or any other natural or
synthetic nanopore is capable of recognizing DNA with single base
resolution.

To address this question, we first investigate the translocation of
poly(dC) with a single position-specific base $A$ substitutions.
Without any loss of generality, we present here only results for
chain length $N=128$, as many of the main conclusions here are not
dependent on the chain length.
Fig. 2 shows the waiting time distribution for
poly(dC$_{31}$dAdC$_{96}$). The waiting time of the monomer $s$ is
defined as the average time between the events that monomer $s$ and
monomer $s+1$ exit the pore~\cite{Luo2,Luo3,Huopaniemi1}. It yields
more detailed information for the translocation process than the
overall translocation time. Compared with the waiting time
distribution for Poly(dC$_{128}$), the waiting times for $s=26 - 36$
are different, where one minimum ($s=28$) and maximum ($s=32$) are
observed.
This result can be easily understood through the following
consideration. The pore of length $L=5$ can accommodate $n_{pore}=6$
monomers on the average. When the monomer $s=32-n_{pore}=26$ is at
the end of the pore ready to exit to the \textit{trans} side, the
monomer corresponding to base $A$ has just entered the pore through
the \textit{cis} side. The presence of the $A$ monomer at this end
of the pore reduces the probability of the backward motion of the
polymer and hence the waiting time starts to decrease starting from
$s=26$ till it reaches a minimum at $s=28$ when the $A$ monomer is
about half way inside the pore, after which the waiting time of
subsequent monomers leaving the pore starts to increase until it
reaches $s=32$ which corresponds to the point where the monomer $A$
is at the end of the pore ready to exit to the \textit{cis} side.
This maximum in waiting time results from the activated nature of
the escape of the monomer $A$ from the pore. After the monomer
corresponding to base $A$ has translocated, the next few monomers
still have longer waiting times compared with those in a homopolymer
Poly(dC$_{128}$). This is due to the fact that the attraction
between the pore and the monomer corresponding to base $A$ induces
more backward events. For other position-specific single base $A$
substitutions, we observed similar results.

\begin{figure}
  \includegraphics*[width=\figurewidth]{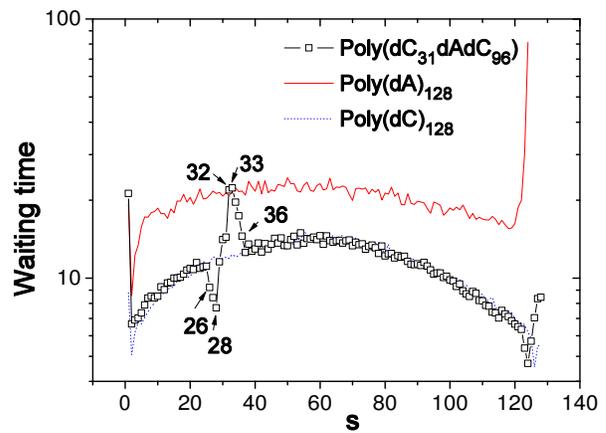}
\caption{The waiting time distribution for poly(dC$_{31}$dAdC$_{96}$).
Here $\epsilon_{pA}=3.0$, $\epsilon_{pC}=1.0$.}
 \label{Fig2}
\end{figure}

\begin{figure}
  \includegraphics*[width=\figurewidth]{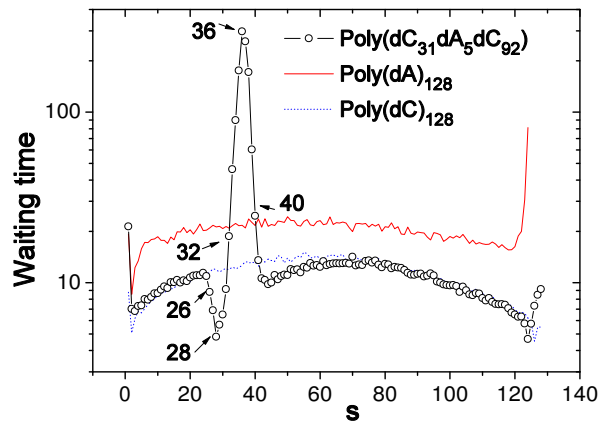}
\caption{The waiting time distribution for poly(dC$_{31}$dA$_5$dC$_{92}$).
Here $\epsilon_{pA}=3.0$, $\epsilon_{pC}=1.0$.}
 \label{Fig3}
\end{figure}

For longer base $A$ sequence substitutions such as
poly(dC$_{31}$dA$_5$dC$_{92}$), the results are shown in Fig. 3.
They can be understood following the same reasoning presented above
for the single $A$ sequence. We still observe one minimum and one
maximum between $s=26 - 36$. As expected, the beginning of the
decrease in waiting time still starts at $s=26$ corresponding to
$s=32-n_{pore}=26$. The maximum now changes to to $s=31+5$ when the
last monomer of type $A$ is at the end of the pore ready to exit to
the \textit{trans} side. For other sequences such as
poly(dC$_{31}$dA$_m$dC$_{97-m}$), the start of the downturn at
$s=26$ and the minimum at $s=28$ remains the same while the maximum
will be shifted to $s=31+m$. Thus, we can conclude that if the
waiting time distribution can be accurately determined, it allows
identification of the sequence $A$ in a straightforward manner.

\subsection{Effect of the different composition on the translocation}

In this section, we present results for the translocation dynamics
for hetero-DNAs with repeat units $A_mC_n$, where $m$ and $n$ are
the length of the bases $A$ and $C$ in a single unit respectively.
We define $m+n$ as the block length of the polymer, and the volume
fraction of the $C$ component as $f_C=n/(m+n)$. In the translocation
dynamics of these hetero-DNAs, it turns out that we need to
distinguish the two different cases where the base $A$ or the base
$C$ first enters the pore, respectively.

We first examine short repeat units with $m+n \le 8$ for the whole
range of volume fractions $f_C$. The results are shown in Fig. 4.
Because the attraction between the pore and the base $A$ is
stronger, the total  translocation time $\tau_A$ for
poly(dA)$_{128}$ ($f_C=0$)is longer than the corresponding value
$\tau_C$ for a poly(dC)$_{128}$ as found in our previous
work~\cite{Luo4}. The solid curve in Fig. 4 represents a simple
average of the translocation time for poly(dA)$_{128}$ and
poly(dC)$_{128}$ given by $\tau(f_C)=\tau_A(1-f_C)+\tau_Cf_C$.
It can be seen that the actual results for the translocation time of
hetero-DNAs with short repeat units are always lower than this
average, indicating a correlation effect between different monomer
blocks. As expected, the translocation time decreases with
increasing $f_C$.
For $m \le 3$ ($f_C \ge 0.6$), the translocation time is almost
independent on the orientation of which base enters the pore first.
This is because the pore with length $L=5$ can accommodate more than
3 monomers. Thus, for $m \le 3$, the translocation times for the
hetero-DNAs are determined largely by the configurations where all
the $A$ monomers are trapped inside the pore, independent of whether
the base $A$ or $C$ enters the pore first. In this regime, the
translocation time can be fitted empirically by the formula
$\tau(f_C)=\tau_Ce^{(1-f_C)}$.

\begin{figure}
  \includegraphics*[width=\figurewidth]{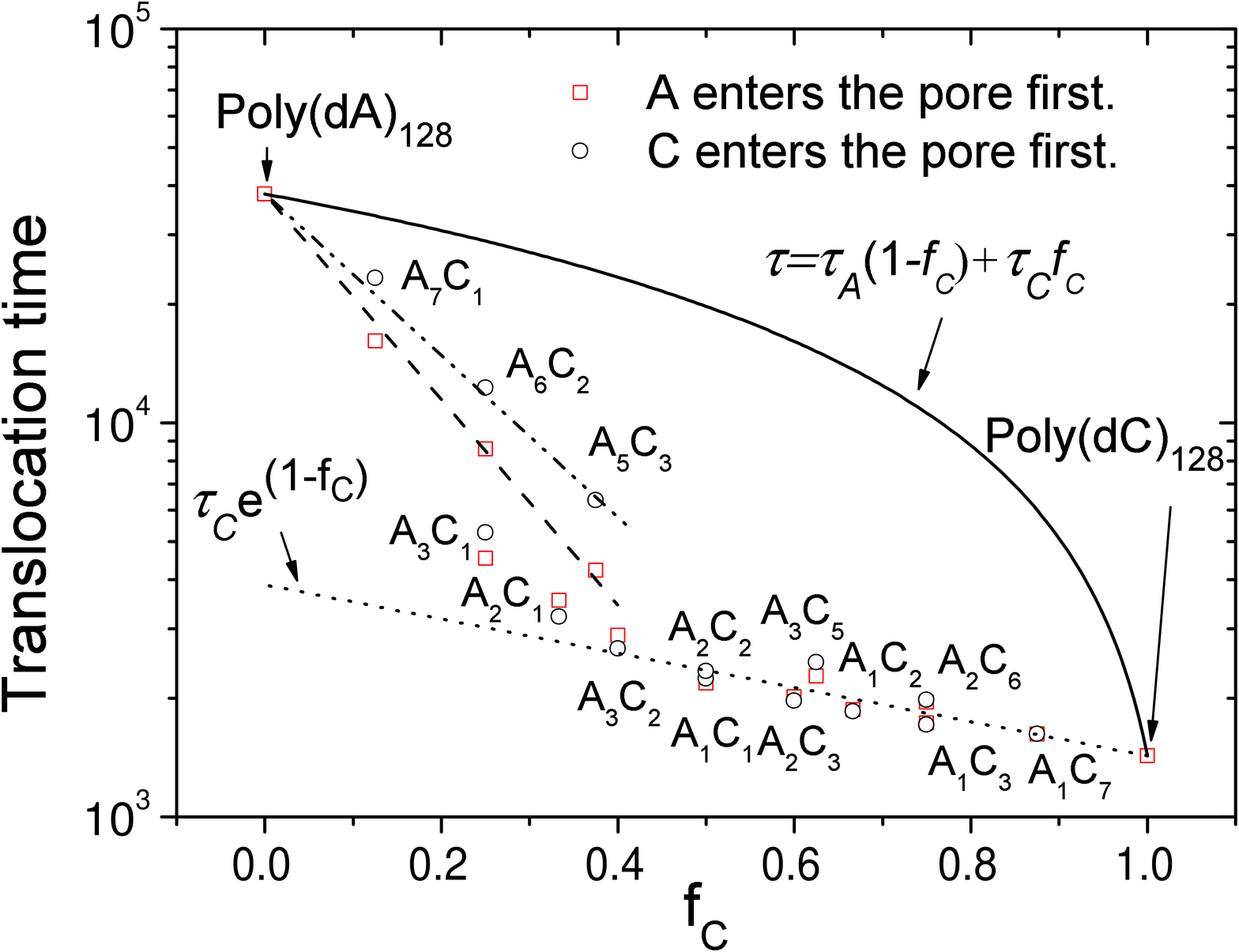}
\caption{Translocation time for hetero-DNAs with repeat units $A_mC_n$
for different values of $m$ and $n$ as a function of the
volume fraction of the base $C$. Here $\epsilon_{pA}=3.0$,
$\epsilon_{pC}=1.0$, and the chain length $N=128$.}
 \label{Fig4}
\end{figure}

For small values of $f_C \le 0.4$ ($m \ge 5$), the translocation
time $\tau$ does depend on the orientation, with $\tau$ being larger
for the case where the base $C$ enters the pore first. This
asymmetry is due to the fact that the translocation time is now
controlled by the last block $A$ exiting the pore. When this is not
followed by a block $C$, the activation barrier for the exit of the
last block is large, leading to a much longer total translocation
time. In this regime, $\tau$ shows faster exponential decay with
increasing $f_C$ when the base $A$ entering the pore first. To
quantify the asymmetry, we define $r$ as the ratio of translocation
times for the case where base $C$ enters the pore first to the case
where $A$ enters the pore first. Fig. 5 shows $r$ for different
sequences. It is clear that the asymmetry increase with the length
of the base $A$ sequence.

\begin{figure}
  \includegraphics*[width=\figurewidth]{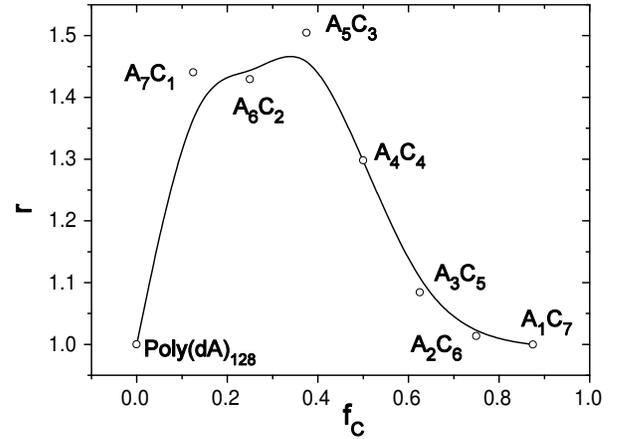}
\caption{$r$ as a function of the different composition for repeat
units $A_mC_n$. Here, we define $r$ as the ratio of translocation
times for base $C$ entering the pore first to base $A$ entering the
pore first. Here, $\epsilon_{pA}=3.0$, $\epsilon_{pC}=1.0$, and the
chain length $N=128$.}
 \label{Fig5}
\end{figure}

Compared to the total translocation time, the waiting time for
individual bases yields more information for the detailed sequence
structure as shown in the last section. To show how the waiting time
can be used to read a periodic sequence, we show the result for
poly(dA$_5$dC$_3$)$_{16}$ in Fig. 6. There are several features
worth noting. First, the waiting time of the last block $A$ is much
longer than the rest, due to the activated nature of the last block
$A$. This effect is more pronounced for $C$ entering the pore first,
since it leads to a much larger activation energy for the exit of
the last block. For the orientation in which $A$ enters the pore
first, both the monomers in the last block $A$ and the last block
$C$ exhibit longer waiting times, but not as strongly as the other
orientation as the corresponding activation energies are smaller.
The orientation of DNA can thus be differentiated in this case. 
Second, the sequence lengths $m=5$ for the base $A$ and $n=3$ for 
the base $C$ are clearly distinguished in comparison with the waiting 
time distributions of their homopolymers.

Moreover, we find that the waiting times for the ordered DNA with
repeat units $A_{m}C_{n}$ exhibit ``fringes'' reminiscent of optical
interference pattern. This is due to the fact that near the end of
exit of any block $A$, the waiting time increases to a maximum
because of the increase in the activation energy. Thus the number of
peaks is exactly equal to $N/(m+n)$, which is the periodicity of 
the sequences (cf. Fig. 6).


\begin{figure}
  \includegraphics*[width=\figurewidth]{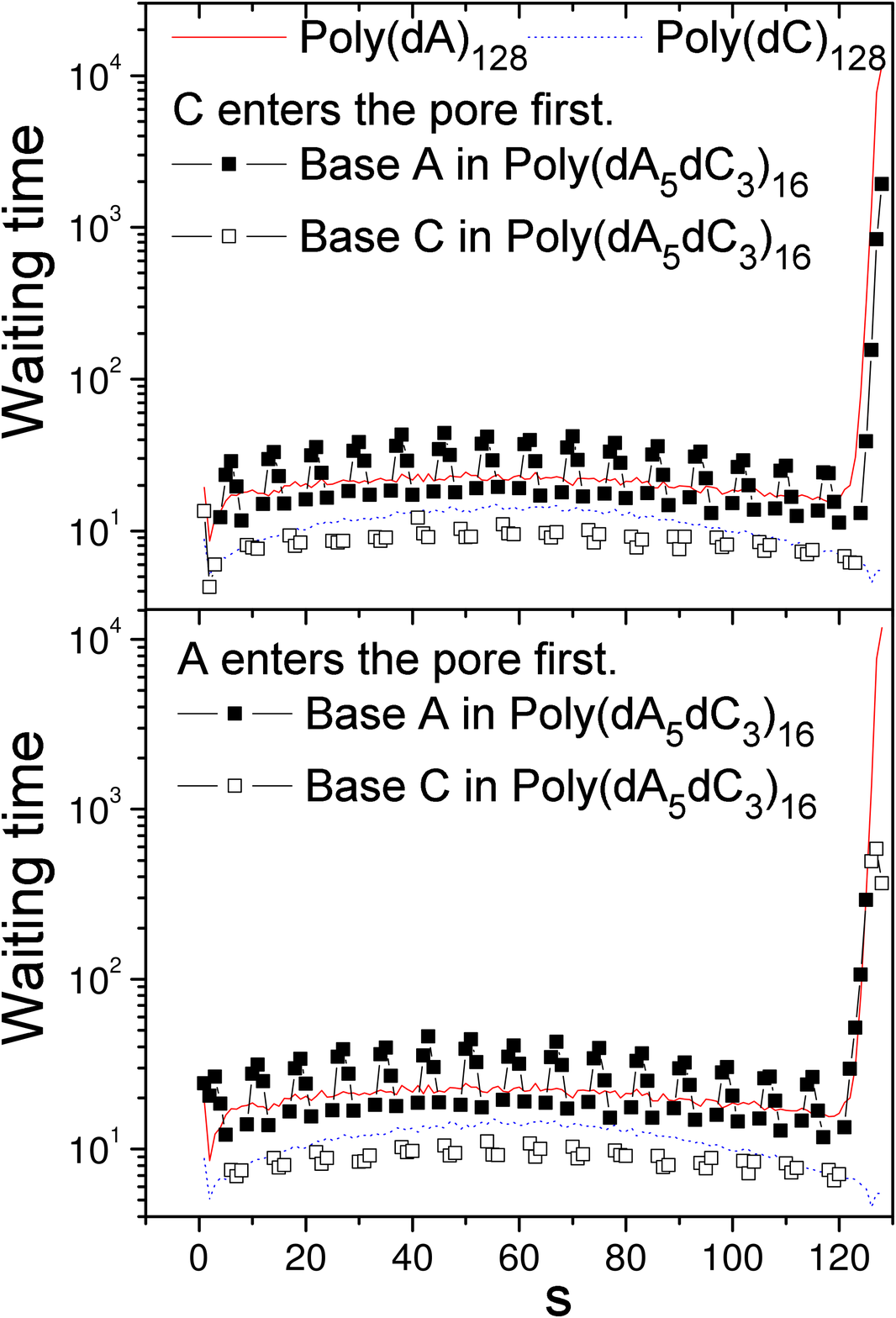}
\caption{The waiting time distribution for poly(dA$_5$dC$_3$)$_{16}$.
Here, $\epsilon_{pA}=3.0$, $\epsilon_{pC}=1.0$,
and the chain length $N=128$.}
 \label{Fig6}
\end{figure}

\subsection{Translocation time as a function of the chain length for different sequences}

Previously, we observed that the translocation time of homopolymers
with pure repulsive interaction with the pore scales as 
$\tau \propto N^{2\nu}$ with the chain length for relatively short 
chains ($N < 200$)~\cite{Luo2,Huopaniemi1}, where $\nu=0.75$ is the Flory
exponent in 2D~\cite{de Gennes, Doi}. 
For heteropolymers consisting of two types of
monomers that have a pure repulsive interaction with the pore and
distinguished only by the magnitude of the driving force that they
experience inside the pore, we find that these scaling properties
remain valid with arbitrary repeat unit~\cite{Luo3}. This could be
easily understood by noting that at a higher level of
coarse-graining, the microstructure of the chain is irrelevant as
far as universal scaling properties are concerned.

When attractive interaction is introduced between the monomers and
the wall of the pore, we have found recently that while the scaling
$\tau \propto N^{2\nu}$ is still valid for poly(dC) with the weaker
attraction, the translocation time $\tau$ shows non-monotonic
behavior as a function of the chain length for poly(dA) due to the
strong attractive force between poly(dA) and the pore~\cite{Luo4}.
This leads to an interesting question on the translocation time
$\tau$ scales with $N$ for different sequences $A_mC_n$ in
heteropolymers considered in this work. In Fig. 7 we plot the
results showing  how the translocation time scales as a function of 
chain length $N$ for different repeat structures 
$A_1C_3$, $A_2C_2$, $A_3C_1$, and $A_6C_2$. 
For $A_3C_1$ and
$A_6C_2$, the translocation time $\tau$ depends on the orientation, but 
the orientation does not affect the qualitative dependence of $\tau$ on
$N$. 

For $A_1C_3$, the scaling exponent is $1.21$ which is close to
$2\nu$. However, with increasing  block length of the base $A$, the
effective scaling exponent decreases. For $A_2C_2$, it is $0.96$ and
it is less than $0.30$ for $A_3C_1$, $A_6C_2$. The novel dependence
on the length of DNA is due to the change from the non-activated
regime for weak attractive or pure repulsive interaction to an
activated regime for strong attractive interaction. It would be
desirable to have measurement made over a larger range of
composition to critically test our predictions for the effective
scaling exponent.

\begin{figure}
  \includegraphics*[width=\figurewidth]{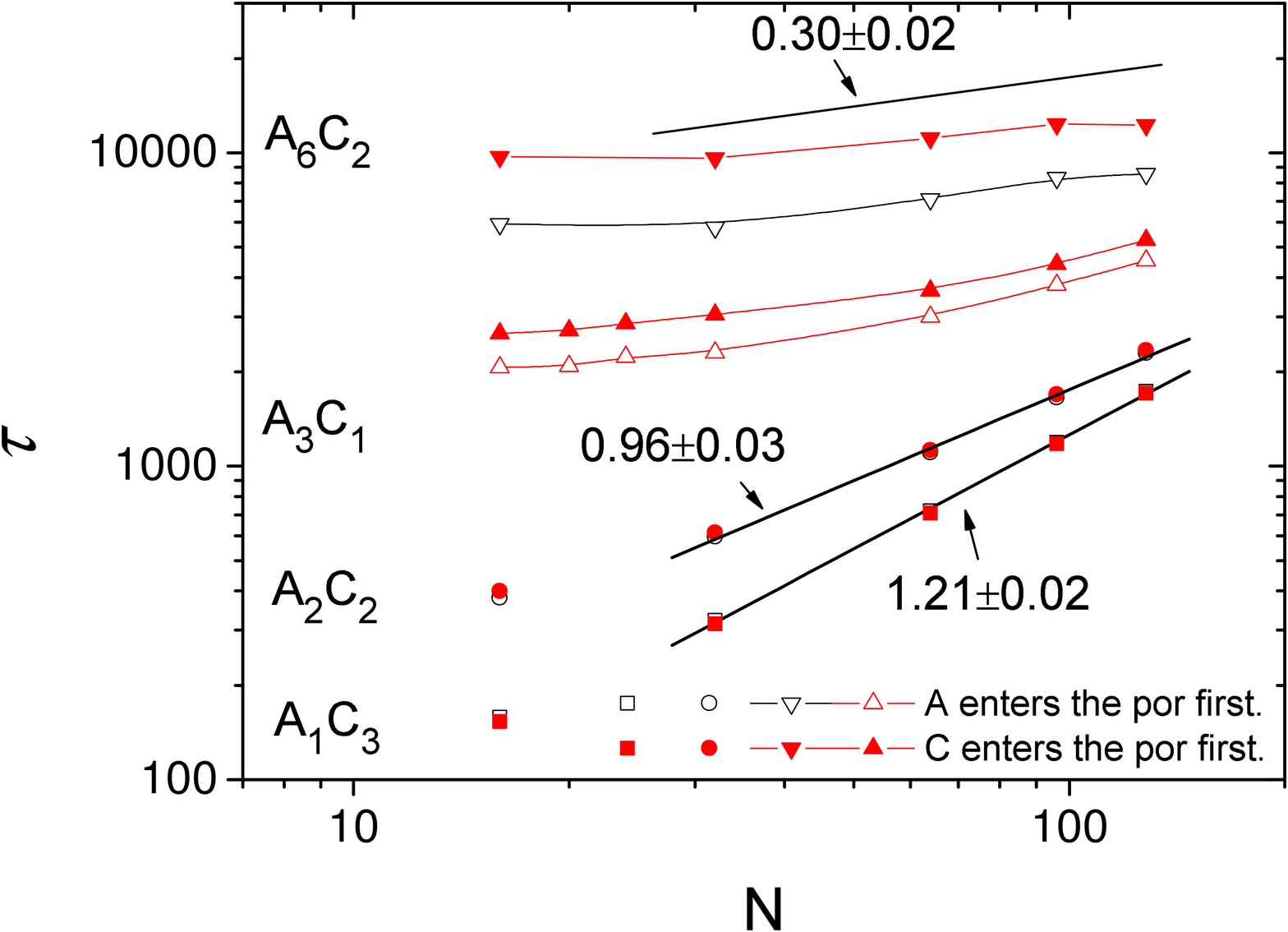}
\caption{Translocation time as a function of the chain length for different sequences.
Here, $\epsilon_{pA}=3.0$, and $\epsilon_{pC}=1.0$}
 \label{Fig7}
\end{figure}

\section{Conclusions} \label{chap-conclusions}

In this work, we have investigated the dynamics of DNA translocation
through a nanopore driven by an external force using 2D Langevin
dynamics simulations with an aim to understand how various aspects of the 
translocation dynamics is affected by a specific sequence. We have   
We have considered coarse-grained models of
hetero-DNAs consisting of two types of bases $A$ and $C$, which are
differentiated by the different base-pore interactions, such as a
strong attractive force between the pore and the base $A$ and a
weaker attractive force between the pore and the base $C$. 
From a series of studies on polymers with sequences $A_mC_n$, we 
find the exponential dependence of the translocation time on $f_C$, 
the volume fraction of the base $C$. For longer sequences with 
$f_C \le 0.5$ the translocation time strongly depends on the orientation 
of DNA, namely the condition which base enters the pore first. 
These results are interpreted according to the waiting times of the individual 
bases. We find that the waiting time displays intriguing fringe patterns that can be
understood from the process of emptying the more attractive subblocks from the pore. 
This also leads to a very simple qualitative understanding of the orientational 
dependence of the translocation time depending on which base enters the pore first. 
Furthermore, based on the waiting time distribution, the sequence lengths $m$, $n$ and 
their periodicity can be distinguished.

One may wonder to what extent our studies are relevant to bringing
this observations into practice and to develop a ``fast sequencing
machine''? Among the most important conclusions here is that the
waiting time distribution of individual monomers can be used to
identify sequences in heteropolymers. It is worth noting in this
context that the ionization potentials for different polynucleotides
are different. Therefore, it is plausible that laser-induced
attractive interactions along with fluorescence spectroscopy could
be used to make a device that will detect either the waiting time or
the number of monomers inside the pore as a function of time.

\begin{acknowledgments}
This work has been supported in part by The Academy of Finland
through its Center of Excellence COMP and TransPoly Consortium
grants. We wish to thank Center for Scientific Computing - CSC Ltd.
for allocation of computer time.
\end{acknowledgments}


\begin{thebibliography}{8}
\bibitem{Tian} P. Tian and G. D. Smith, \textit{J. Chem. Phys.} {\bf 119}, 11475 (2003).
We note that the results here contradict both experiments as well as
other simulation studies.
\bibitem{Note}  $W$ is measured from the center of the wall particles. For a value of $W=3\sigma$,
the actual separation between the outermost boundaries of the two walls is $2\sigma$.
\end{thebibliography}
\end{document}